\documentclass[10pt,conference]{IEEEtran}
\usepackage{graphicx}
\usepackage{subcaption}
\usepackage{float}
\usepackage{booktabs}
\usepackage{cite}
\usepackage{amsmath,amssymb,amsfonts}
\usepackage{siunitx}
\usepackage{braket}
\usepackage{listings}
\usepackage{titlesec}
\usepackage{inconsolata}
\usepackage{pifont}
\usepackage{minted}

\titlespacing*{\subsubsection}
{0pt}{1.2ex plus 0.5ex minus 0.2ex}{0.8ex}
\usepackage{xcolor}         
 \usepackage[colorlinks]{hyperref}

\usepackage{mathtools}

\title{Retromorphic Testing of Quantum Compiler Passes}

\begin{document}
 
\author{\IEEEauthorblockN{Mushahid Khan and Olivia Di Matteo }
\IEEEauthorblockA{\textit{ Electrical and Computer Engineering and Stewart Blusson Quantum Matter Institute} \\
\textit{The University of British Columbia} \\
Vancouver, Canada \\
mkhan103@student.ubc.ca, { olivia}@ece.ubc.ca}
}

\maketitle

\begin{abstract}

Quantum compilers play a critical role in transforming high-level quantum programs into optimized, hardware-compatible circuits. However, verifying the correctness of compiler passes remains challenging, as determining the expected output of large, deeply entangled quantum circuits is computationally intractable. This challenge is further amplified when compiler passes modify already complex circuit structures, making manual validation of transformed circuits impractical. 

In this work, we perform a systematic analysis of unit tests for quantum compiler passes in four quantum programming frameworks (PennyLane, Qiskit, Cirq, and pytket). Our findings indicate validation is dominated by program-content and program-metric assertions, and test circuits are generally small and shallow. Motivated by these observations, we introduce a testing methodology for automated validation of quantum compiler passes based on retromorphic testing and principles from the Hadamard test. This methodology analyzes a compiler pass, test circuit, and expected pass behavior to verify semantic preservation and intended structural modifications. 

We implement our methods in a framework, RetroQ, and apply it to compiler passes in PennyLane and Qiskit. Experimental evaluation reproduced several existing bugs as well as uncovered previously undetected defects, such as flawed symbolic parameter handling, incorrect commutation logic, failure to recognize self-adjointness of gates, and runtime crashes. These findings highlight the need for compiler-pass-specific testing methodologies to improve the reliability of the evolving quantum software stack.

\end{abstract}

\section{Introduction}
Quantum computing represents a paradigm shift in computational power, promising to solve problems in applications such as cryptography and materials science that are intractable for classical systems~\cite{mavroeidis2018impact, lordi2021advances}. However, the path from a high-level quantum algorithm to a physical result on hardware is complex and requires dedicated quantum software engineering methods~\cite{stirbu2023full, 10.1145/3712002, zhao2021quantumsoftwareengineeringlandscapes}. A key part of this quantum software stack is the quantum compiler, which transforms high-level programs into optimized, hardware-compatible circuits through a series of passes or transformations. 
Compilation passes reduce target metrics, often circuit depth and gate count, with optimizations such as unitary gate unification (which merges operations) and gate cancellation (which removes gates followed by their adjoints), while ensuring compatibility with specific hardware constraints through qubit mapping and gate scheduling.

Compiler passes are generally equivalence-preserving, meaning the output remains unchanged despite structural optimizations. Validation can be performed with equivalence checking, a low-level method that involves comparing matrix representations of two circuits~\cite{chareton2022formalmethodsquantumprograms, Wille_2024}. This is distinct from higher-level formal verification methods, such as Quantum Hoare Logic~\cite{ying2019toward}, which demonstrate that a circuit meets its intended mathematical specifications. Both approaches, however, face significant hurdles~\cite{chareton2022formalmethodsquantumprograms, ying2019toward, quetschlich2025experience, govindankutty2026review}, such as the exponential growth of the state space, making it extremely difficult to classically simulate or mathematically model the behavior of large quantum circuits. As a result, developers must pivot to more accessible validation and debugging techniques, such as unit testing, to verify compiler passes. However, unit testing is hindered by the test oracle problem, a long-standing bottleneck in classical software and compiler testing that arises because of system complexity and a lack of clear specifications.



The test oracle problem is the challenge of distinguishing a system's valid, desired output from incorrect behavior for any given input. This problem is complicated in quantum programming due to limited human intuition regarding features such as superposition and entanglement~\cite{quetschlich2025experience}. While unit tests may verify simple cases, they may fail to provide coverage for large, entangled circuits for which determining the expected output is computationally intractable. This difficulty is often heightened by the compound nature of quantum bugs, which typically occur at the intersection of multiple categories~\cite{quetschlich2025experience}.

Black-box testing techniques~\cite{nidhra2012black} provide a natural way to address the test oracle problem by evaluating program behavior without knowledge of the internal structure. This raises the question of whether it can be used to reliably verify the correctness of quantum compiler passes when manual oracles are computationally intractable for large circuits and formal verification remains complex. To address this problem, we formulate the following research questions (RQ\#):
\begin{itemize}
\item \textbf{RQ1:} How are quantum compiler passes currently tested and validated, and to what extent are existing approaches sufficient for ensuring their correctness?
\item \textbf{RQ2:} How can black-box testing techniques verify quantum compiler passes without explicit computation of the expected circuit output?
\item \textbf{RQ3:} What types of compiler pass errors are not captured by existing testing and verification approaches?
\end{itemize}

To answer these research questions, we developed a testing framework called RetroQ. RetroQ leverages retromorphic testing~\cite{yu2023retromorphic}, a black-box methodology inspired by inverse function reasoning, the principles underlying the Hadamard test~\cite{h_test}, and knowledge of the expected behavior of the compiler pass to systematically validate compiler correctness. Although implemented with support for testing the PennyLane~\cite{bergholm2022pennylaneautomaticdifferentiationhybrid} and Qiskit~\cite{javadiabhari2024quantumcomputingqiskit} frameworks, the proposed methodology can be applied to compiler transformations in any framework. It is also generalizable beyond compiler passes to other classes of quantum software: because quantum operations are unitary and reversible, retromorphic testing can be used to verify quantum arithmetic modules, uncomputation subroutines, or encoding circuits in quantum error correction. 

RetroQ allows users to input a set of compiler passes along with a target quantum function and expected behavior of each compiler pass, enabling automated evaluation of each compiler pass’s behavior without manual derivation of expected outputs. 
Experimental evaluation revealed correctness issues in quantum compiler passes in PennyLane and Qiskit, such as runtime crashes during inverse cancellations on arbitrary unitary operators, failure to remove identity-equivalent gates when utilizing symbolic parameters, and incorrect commutation logic that led to invalid gate removals. Also, our results expose structural limitations in pattern-matching optimizations for self-adjoint operators and underscore the importance of considering global phase in a controlled context. The complexity of RetroQ scales with the target function and the compiler pass applied to it, with overhead no worse than that expected from applying the pass itself.


The remainder of this paper is organized as follows. Section \ref{sec:background} reviews the relevant background and literature. Section \ref{sec:unittest} analyzes existing unit testing approaches in PennyLane, Qiskit, Cirq~\cite{cirq}, and pytket~\cite{pytket_docs}. Section \ref{sec:ava} presents RetroQ and its two-stage verification approach. Section \ref{sec:res} reports the evaluation results and discovered compiler bugs. Finally, Section \ref{sec:conclusion} concludes the paper and outlines future work.

\section{Background and Related Work}
\label{sec:background}
We assume familiarity with fundamental quantum concepts (qubits, gates, adjoints, unitary matrices, measurements, expectation values) and metrics for quantum circuits (circuit depth and gate count)~\cite{nielsen2010quantum}. This section introduces retromorphic testing, quantum software testing concepts, and related work that provide necessary background for our methodology. 


\subsection{Retromorphic Testing}
The test oracle problem is a significant challenge in software testing, as determining output correctness without a predefined standard or manual intervention is difficult and time-consuming~\cite{yu2023retromorphic}.
Black-box testing techniques are one approach to the test oracle problem~\cite{yu2023retromorphic}. In black-box testing, the tester does not require knowledge of the program’s internal implementation details~\cite{nidhra2012black}, evaluating the system solely through its input-output behavior. Three black-box testing techniques are differential testing~\cite{mckeeman1998differential}, metamorphic testing~\cite{chen2020metamorphictestingnewapproach}, and retromorphic testing~\cite{yu2023retromorphic}. 

Metamorphic testing~\cite{chen2020metamorphictestingnewapproach} defines expected relationships between multiple executions of a program under specific input transformations; violations of these relationships may indicate the presence of defects. In contrast, differential testing~\cite{mckeeman1998differential} compares the outputs of multiple versions or implementations of the same system when given identical inputs, with inconsistencies revealing potential bugs.

Retromorphic testing~\cite{yu2023retromorphic}, which forms the basis of our work, is motivated by the relationship between a function and its inverse. It utilizes a dual-program structure consisting of a forward program ($f$) and its inverse ($f^{-1}$). Depending on the system under test, either program may serve as the primary program while the other acts as the auxiliary program to reverse the transformation and restore the original input. Bugs can be identified by examining the relationship between the initial input and the final output,  without requiring knowledge of how $f$ is actually implemented. Retromorphic testing is the approach we use to address \textbf{RQ2}.

To illustrate this idea, consider a simple forward program $f(x)$ that increments an integer input by one, with its inverse  $f^{-1}(x)$ decrementing the input by one. Starting with $x=3$, the forward program produces 4, after which the inverse returns the value of $3$. Since the final output matches the original input, the transformation is considered correct.

\subsection{Quantum Software Testing}
Quantum software testing can be divided into two parts: the testing of quantum programs and the testing of quantum platforms~\cite{paltenghi2024surveytestinganalysisquantum}.
Testing at the program level focuses on verifying application-specific code. Testing  techniques for quantum programs include combinatorial testing~\cite{9724888}, search-based test generation~\cite{wang2021generating}, runtime verifications using statistical, auxiliary-  and projection-based assertions~\cite{huang2019statistical, li2019proq, 10.1145/3373376.3378488}, and static analysis methods based on formal verification, abstraction, and pattern detection~\cite{10189127}.

In this work, we focus on platform-level testing. Platform-level testing targets three main components: a quantum programming language to express programs; a compiler to transform the program before execution; and an execution environment to run programs on simulators or quantum hardware~\cite{paltenghi2024surveytestinganalysisquantum}. Platform-level testing targets the general equivalence of program variants or the stability of the entire software stack. Examples of platform-level testing include QDiff~\cite{wang2021qdiff}, MorphQ~\cite{paltenghi2023morphq}, and MorphQ++~\cite{kitt2024morphq++}.

QDiff~\cite{wang2021qdiff} targets the whole stack by utilizing differential testing across different quantum software frameworks. It identifies instabilities by generating logically equivalent witness programs from a seed through the application of seven gate transformation rules. It then compares their measurement distributions with a reference distribution across a software stack using the Kolmogorov-Smirnov test~\cite{berger2014kolmogorov} and cross entropy~\cite{Shore1980}. This methodology can be computationally expensive as it requires multiple program variants to identify a bug rather than an internal, self-contained source of truth.

 MorphQ~\cite{paltenghi2023morphq} and its successor, MorphQ++~\cite{kitt2024morphq++}, apply metamorphic testing to the entire stack of a single platform (Qiskit). MorphQ(++) generates syntactically correct program pairs through automated transformations such as changing the optimization level. These transformations force the entire platform infrastructure to process the program. Although these tools are based on metamorphic testing, they did not identify faults by comparing output relationships; instead, they identified faults through unexpected crashes~\cite{kitt2024morphq++}.

QDiff, MorphQ, and MorphQ++ focus on testing single platforms across different configurations. Multi-platform testing is addressed by QITE~\cite{paltenghi2025qite}. QITE is a general framework that leverages quantum assembly language as a common representation to bridge frameworks, and uses a crash oracle and equivalence oracle to detect failures across platforms.


The aforementioned methodologies are holistic because they aggregate errors from all layers (program generation, optimization, mapping, and hardware noise) into one final measurement or a system-wide crash, rather than isolating specific internal compiler transformations.

In contrast, some frameworks focus on formally verifying individual internal compiler transformations. 
VOQC (Verified Optimizer for Quantum Circuits) uses the Coq proof assistant~\cite{coq_development_team_2019} to verify optimizers for quantum circuits. It utilizes a symbolic intermediate representation, which allows it to reason about circuits with an arbitrary number of qubits. However, this approach requires specialized expertise and time investment, as manual proofs can exceed the actual code size by more than an order of magnitude~\cite{Hietala_2021}. 

Giallar~\cite{tao2022giallarpushbuttonverificationqiskit} uses SMT (Satisfiability Modulo Theories) solvers to verify that Qiskit compiler passes preserve quantum semantics by extracting invariants from loop templates and using pre-verified rewriting rules. However, a reliance on a fixed set of loop templates and incomplete rewrite rules restricts its ability to verify future compiler passes with unique loop patterns.

\section{Quantum Computing Framework Unit Testing}
\label{sec:unittest}

To answer \textbf{RQ1} (How are quantum compiler passes currently tested
and validated, and to what extent are existing approaches
sufficient for ensuring their correctness?), we analyzed assertion statements used in existing unit tests for compiler passes in PennyLane, Cirq, Qiskit, and pytket. Specifically, we examined 113 test files for PennyLane, 31 for Qiskit, 55 for Cirq, and 1 for pytket\footnote{Only a single Python test file in pytket includes tests for compiler passes.}. 

To systematically extract these assertions, we implemented a static analysis pipeline that parses each test file into its Abstract Syntax Tree (AST) representation using Python’s built-in \texttt{ast} module. We then traversed the AST to identify assertion nodes, and grouped assertions on a per-file basis. We manually examined each assertion within its corresponding test file to understand the intent of the unit test, and categorized assertions based on the specific circuit or framework properties being validated.
The categories used are as follows:
\begin{itemize}
  \item \emph{Program Metrics}: Quantitative quantum circuit characteristics, such as gate count, circuit depth, and qubit count.
  \item \emph{Program Contents}: The presence, absence, or ordering of contents in a quantum circuit, such as gates or qubits.
  \item \emph{Internal Logic and Data Structure}: Properties of internal compiler components such as names, internal data structures, helper functions, and other framework-specific implementation details.
   
\item \emph{Matrix Equivalence} (of quantum circuits, or individual operations)

\item \emph{Output Equivalence}: The original and compiled quantum circuits produce identical outputs, such as state vectors, probability distributions, or expectation values.

\item  \emph{Program Execution Details}: Quantum circuit execution related information, such as the number of shots.

\item \emph{Output Property Equivalence} (such as statistical characteristics or structural properties of state vectors)
\item \emph{Exception Handling}: Whether expected exceptions are raised, or errors are reported appropriately.
\item \emph{Gradient Equivalence}: Verification that the transformed quantum program produces correct gradients, ensuring the accuracy of differentiable quantum computations.

  \end{itemize}

Our findings, shown in~\autoref{tab:my_table}, reveal that validation of compiler passes is a multi-faceted approach that spans structural integrity, internal logic, and mathematical accuracy. 
Since compiler passes frequently rewrite or optimize quantum circuits, verifying the circuit contents is the dominant testing strategy across frameworks.
However, testing practices are not uniform, and developers prioritize different correctness properties depending on the framework.

Pytket differs substantially from the other frameworks, with 50.88\% of the assertions focused on Program Metrics. This indicates a stronger emphasis on validating optimization-related properties such as depth and gate count. 

Cirq exhibits a comparatively high proportion of Matrix Equivalence assertions, indicating that mathematical correctness and equivalence of transformed quantum operations are important validation concerns. PennyLane is the only framework that tests gradient equivalence, reflecting its support for differentiable quantum programming.

Additionally, categories such as Program Execution Details and Output Property Equivalence appear infrequently across frameworks, suggesting that compiler pass testing is generally focused on structural and semantic correctness rather than execution configurations or statistical output characteristics. 

\begin{figure}[ht]
         {\includegraphics[width=0.45\textwidth]{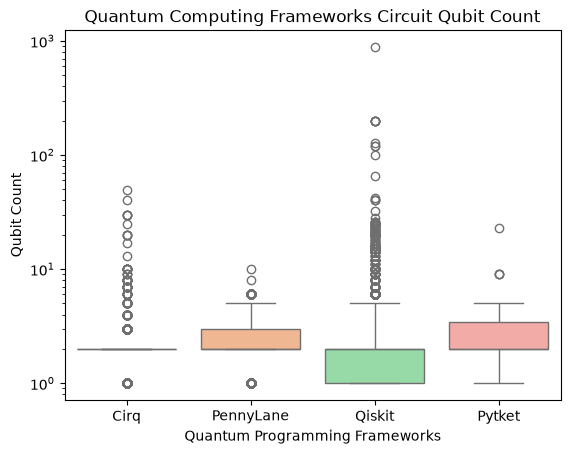}}
        \caption{Qubit count distribution of unit test circuits. }
        \label{fig:cc}
\end{figure}

\begin{figure}[ht]
         {\includegraphics[width=0.45\textwidth]{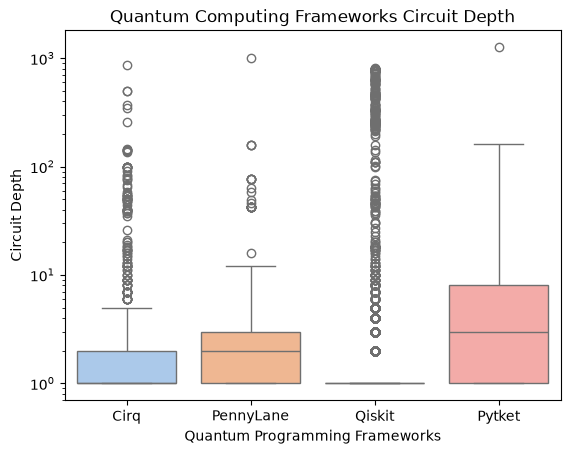}}
        \caption{Depth distribution of unit test circuits.}
        \label{fig:cd}
\end{figure}

We also analyzed the quantum circuits used in these unit tests, focusing on their circuit depth and qubit counts, as these capture the scale of quantum programs used during testing. To collect these metrics, we implemented framework-specific Python functions to extract and record the circuit depth and qubit count of each test circuit. We instrumented the unit tests to invoke these functions whenever a circuit was constructed and ready to be input to a compiler pass. We then executed the test suites for each framework and automatically recorded the resulting metrics. In total, we examined 1,518 circuits for PennyLane, 7,559 for Qiskit, 1,682 for Cirq, and 119 for pytket. Box plots of qubit count and circuit depth are presented in \autoref{fig:cc} and \autoref{fig:cd}, respectively.

Most quantum circuits in unit tests use a small number of qubits. Qiskit exhibits the widest qubit count distribution and the largest number of circuits overall. Qiskit also has several extreme outliers with approximately 100, 200, and nearly 900 qubits. These outliers stem from targeted scalability stress tests for the \texttt{StarPrerouting} and \texttt{LightCone} transpiler passes, fake backend simulations mirroring the 127-qubit \texttt{ibm\_sherbrooke} architecture, and large-scale circuits generated directly from 20$\times$20 hexagonal lattice graphs.
Circuits in Cirq, pytket, and PennyLane, by contrast, are concentrated below 50 qubits with relatively few high-qubit test cases.




Across all frameworks, circuit depth is heavily concentrated in the shallow regime, with the majority of circuits falling between depth 1-10. Cirq, PennyLane, and Qiskit are especially dominated by very shallow circuits ($ \approx84-94\%$ in the 1-5 range and $\approx95\%$ within 1-10), while pytket shows a comparatively broader spread, with 66.10\% in 1-5 and 81.36\% within 1-10. Qiskit exhibits some high-depth outliers, and Cirq shows a long-tailed distribution of deeper circuits. The highest outlier in PennyLane results from a test that evaluated whether deeply nested pairs are canceled using the \texttt{cancel\_inverses} pass. For pytket, the outlier came from a test that applied Peephole Optimization to a circuit loaded from a JSON file.

In summary, compiler passes are currently validated through assertion-based unit testing that prioritizes structural integrity and mathematical accuracy, primarily by verifying circuit contents and program metrics. While validation strategies vary by framework, current methods are largely limited to smaller circuit depths and qubit counts. The reliance on simple circuits suggests that current methodologies may miss bugs that might only manifest in larger circuits. However, determining the expected output for large, deeply entangled quantum circuits is computationally intractable.

\begin{table}[htbp]
\caption{Categorization of assert statements in unit tests for compilation passes. }
\label{tab:my_table}

\label{tab:quantum_frameworks_data_percent}
\small
\centering
\setlength{\tabcolsep}{4pt} 
\begin{tabular}{|l|c|c|c|c|}
\hline
\textbf{Groups } & \textbf{PennyLane} & \textbf{Qiskit} & \textbf{Cirq} & \textbf{pytket} \\ \hline

Program Metrics & 16.08\% & 14.39\% & 3.70\% & 50.88\% \\ \hline

Program Contents & 42.85\% & 49.56\% & 62.24\% & 39.93\% \\ \hline

\begin{tabular}[c]{@{}l@{}}
Internal Logic \\
and Data Structure
\end{tabular}
& 9.62\% & 14.65\% & 5.41\% & 2.47\% \\ \hline

Matrix Equivalence & 3.10\% & 8.81
\% & 21.35\% & 1.41\% \\ \hline

Output Equivalence & 10.81\% & 0.06\% & 2.37\% & 3.89\% \\ \hline

\begin{tabular}[c]{@{}l@{}}
Program Execution \\
Details
\end{tabular}
& 1.32\% & 8.09\% & 0.00\% & 0.00\% \\ \hline

\begin{tabular}[c]{@{}l@{}}
Output Property \\
Equivalence
\end{tabular}
& 5.08\% & 0.00\% & 0.00\% & 0.35\% \\ \hline

Exception Handling & 4.28\% & 4.44\% & 4.93\% & 1.06\% \\ \hline

Gradient Equivalence & 6.85\% & 0.00\% & 0.00\% & 0.00\% \\ \hline

\end{tabular}
\end{table}

\section{Retromorphic Quantum Compiler Testing (RetroQ)}
\label{sec:ava} 
To address the limitations of current unit testing methodologies and the computational intractability of verifying large-scale, deeply entangled quantum circuits, we introduce RetroQ. RetroQ is a testing framework designed to systematically validate the correctness of quantum compiler passes and answer \textbf{RQ2} (How can black-box testing techniques verify quantum compiler passes without explicit computation of the expected circuit output?).
RetroQ operates by generating transformed versions of a quantum circuit using a compiler pass and then verifying whether the semantic behavior and structural properties are preserved. It achieves this by combining retromorphic testing with the Hadamard test~\cite{h_test} to compare original and transformed circuits without requiring manual specification of expected outputs. RetroQ is available on GitHub \cite{retroq2026}.
 
Let $F$ be a target quantum circuit and let $\zeta$ denote a compiler pass. Mathematically, the application of the compiler pass produces a transformed circuit, $F'=\zeta(F)$.
Two quantum circuits are considered equivalent, denoted as $F \equiv F'$, if they produce identical outcomes for all input states, up to numerical tolerance and global phases. This notion of equivalence forms the basis for semantic correctness in RetroQ. 

RetroQ models compiler pass correctness as the satisfaction of two predicates: one over semantic information and one over structural properties. A semantic preservation predicate is defined as:
$\mathcal{F}_{sem}(F, F')$, which is equal to true if $ F \equiv F'$ and false otherwise.
       
Predicates over structural properties are represented as predicates over circuit metrics. Let $g(F)$ denote the amount of (a specific type of) quantum operations of the circuit $F$. RetroQ currently supports predicates of the form:
$\mathcal{F}_{gate}(F, F') \in \{g(F') < g(F),~g(F') = g(F),~g(F') > g(F)\}$.
More generally, any circuit metric $m(\cdot)$, such as depth or qubit count, may be constrained through predicates of the form $\mathcal{F}_m(F, F'):= m(F') \ast m(F)$, where $\ast \in \{ <,~ 
\leq,~ = , ~ \geq, ~>\}$.


RetroQ evaluates whether the transformed circuit satisfies the semantic information and structural property predicates using a two-stage verification process: an adjoint identity check and a controlled adjoint identity check.


\subsection{Stage 1: Adjoint Identity Check}
\begin{figure}[ht]
         {\includegraphics[width=0.45\textwidth]{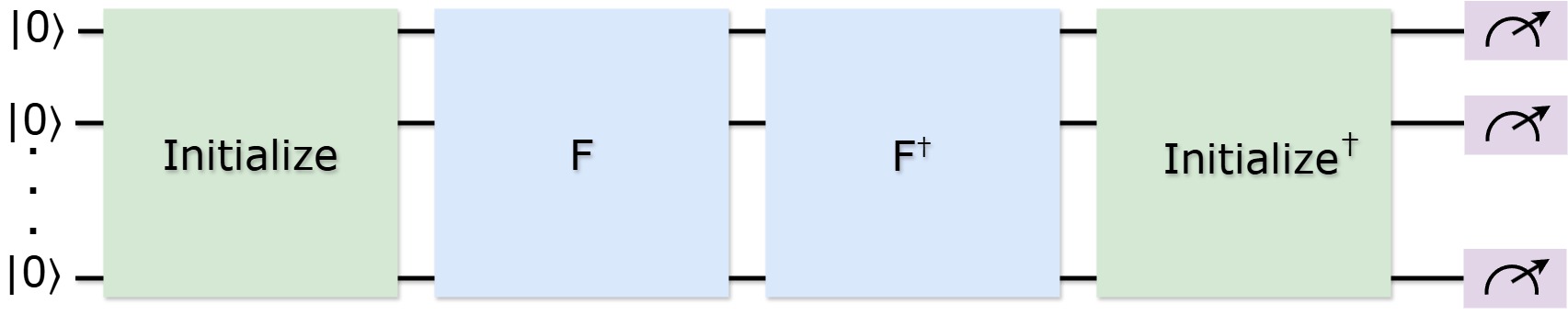}}
        \caption{Stage 1 baseline circuit.}
        \label{fig:baseline_figure1}
\end{figure}

\begin{figure}[htbp]
    \centering
        \begin{subfigure}{\linewidth}
        \centering
        \includegraphics[width=0.90\linewidth]{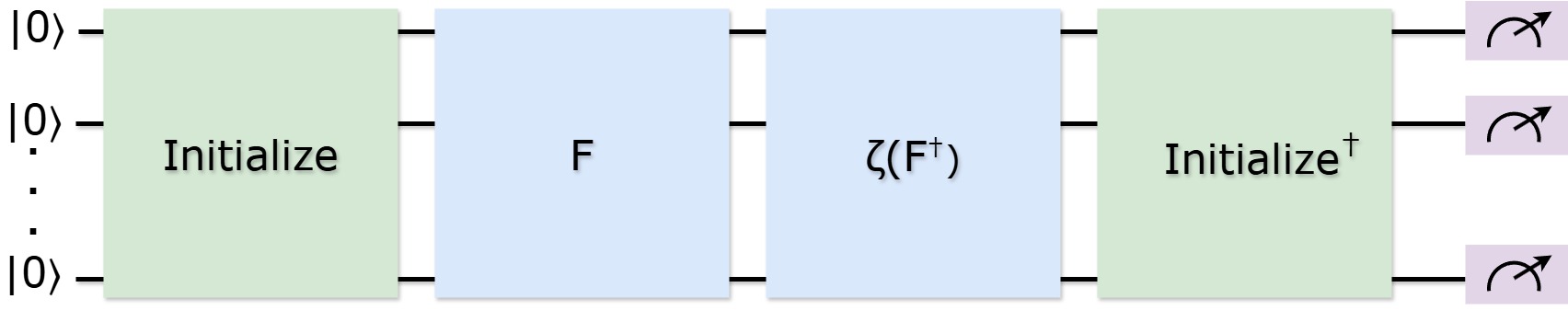} 
        \caption{}
        \label{fig:s1top_image}
    \end{subfigure}
    \begin{subfigure}{\linewidth}
        \centering
        \includegraphics[width=0.90\linewidth]{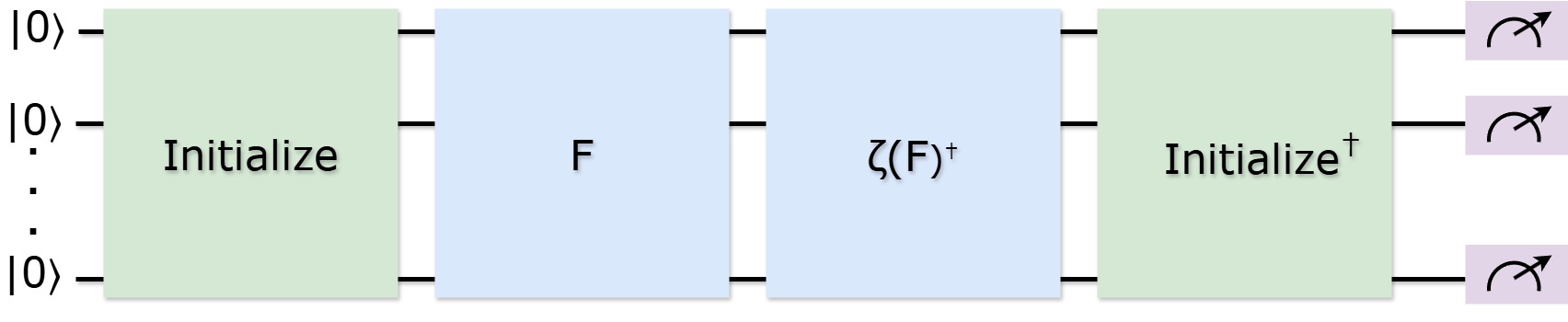} 
        \caption{}
        \label{fig:s1bottom_image}
    \end{subfigure}
    
    \vfill 
    

    \caption{Stage 1 variant quantum circuits. $\zeta$ is a compiler pass.}
    \label{fig:stage1}
\end{figure}

In Stage 1, RetroQ implements a basic retromorphic identity check. RetroQ first generates a baseline quantum circuit (\autoref{fig:baseline_figure1}) that prepares the qubits in a known state using an initialization operator, then applies $F$, followed by the adjoint of $F$, and finally the adjoint of the initialization operator. In this baseline setting, all qubits are expected to be in the $\ket{0}$ state upon measurement. RetroQ then generates two transformed versions of the original circuit that represent alternative ways of combining the target circuit, its adjoint, and the compiler pass: $\zeta(F^{\dag})$ and $\zeta(F)^{\dag}$. These two variants can be seen in \autoref{fig:stage1}.


Testing compiler passes on adjoints of operations is important because adjoints arise in several common scenarios, including uncomputation and reversible subroutine constructions. As we will see in \autoref{sec:res}, the choice of how adjoint operations are represented can have an impact on how compiler passes behave.


The current implementation of RetroQ uses the identity as the default initialization function. This design choice reflects the current scope of the framework, which focuses on verifying compiler passes that depend only on the sequence and structure of quantum operations.
In principle, RetroQ can employ any initialization routine. Alternative choices may expose different compiler pass behaviors if a pass is sensitive to the input state. 

For each program in Stage 1, RetroQ records the semantic information, by verifying whether each qubit is returned to $\ket{0}$ (same as the baseline circuit), and structural information. The semantic and structural information are then validated against the corresponding baseline values. Deviations from the baseline indicate that one or more verification predicates were violated.

RetroQ's automatic derivation and checking of these expected properties provides a test oracle for compiler pass validation. This removes the need for users to manually compute expected outputs for large quantum circuits, enabling more scalable verification of compiler transformations.

\subsection{Stage 2: Controlled Adjoint Identity Check}
The Hadamard test~\cite{h_test} is a quantum computing method used to estimate the overlap between quantum states. 
Specifically, it can be used to estimate the real part or imaginary part of the expectation value $\bra{\psi} U \ket{\psi}$, where $\ket{\psi}$ is a quantum state and $U$ is a unitary operator. 

The test introduces an auxiliary qubit initialized in the $\ket{0}$ state, to which a Hadamard gate is applied, creating a superposition. A controlled-$U$ operation is then performed. A second Hadamard gate is applied before measuring the auxiliary qubit. Once the auxiliary qubit is measured, the probability of measuring state $\ket{0}$ will be $\frac{1}{2} \cdot (1 + \hbox{Re} \bra{\psi}U\ket{\psi})$ and the probability of measuring the state  $\ket{1}$ will be $\frac{1}{2} \cdot (1 - \hbox{Re} \bra{\psi}U\ket{\psi})$, where $\hbox{Re}$ refers to the real part of a complex number.
With these probabilities, the expectation value $\langle Z\rangle$ is equal to $\hbox{Re}\bra{\psi}U\ket{\psi}$. If $U$ is the Identity operator, the expectation value should be $+1$, indicating that the auxiliary qubit is back in the $\ket{0}$ state.

\begin{figure}[ht]
         {\includegraphics[width=0.42\textwidth]{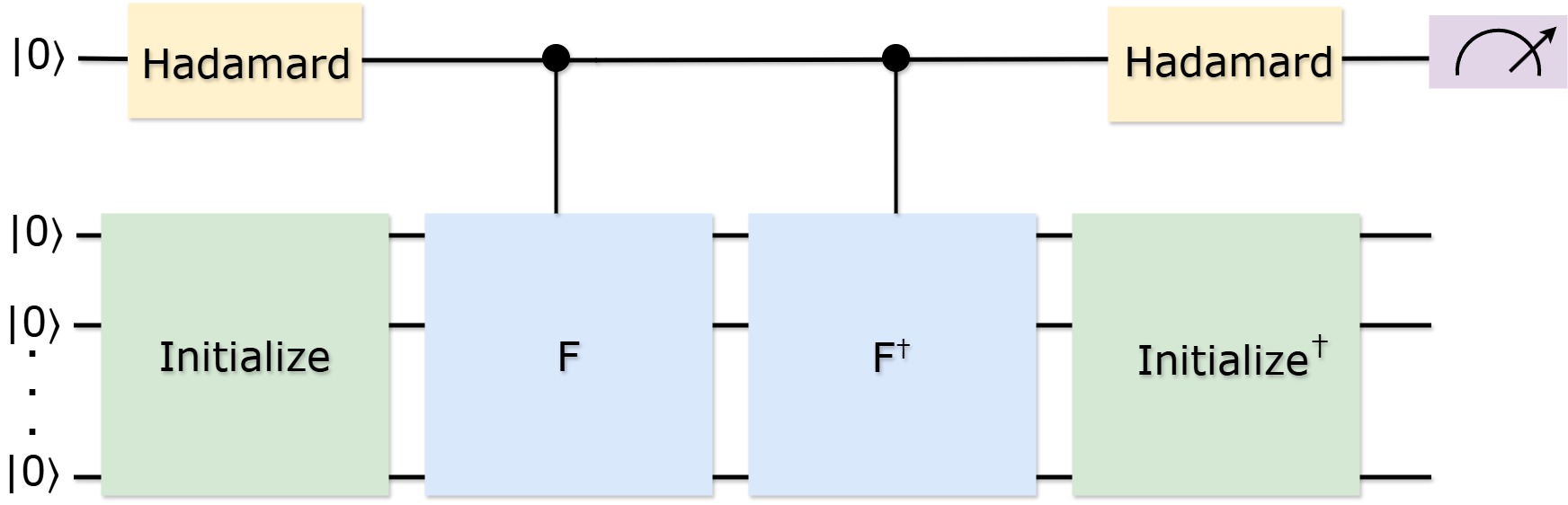}}
        \caption{Stage 2 baseline circuit.}
        \label{fig:stage2_baseline}
\end{figure}

In Stage 2, RetroQ evaluates whether the transformed circuit satisfies the semantic information and structural properties predicates when the compiler pass is used within a controlled operation. To achieve this, the framework first generates the baseline untransformed quantum circuit resembling the Hadamard test circuit, as shown in \autoref{fig:stage2_baseline}. When the auxiliary qubit is measured at the end, it is expected to be in the state $\ket{0}$, yielding an expectation value of $+1$. 
 
\begin{figure}[htbp]
    \centering

    \begin{subfigure}{\linewidth}
        \centering
        \includegraphics[ width=0.90\linewidth]{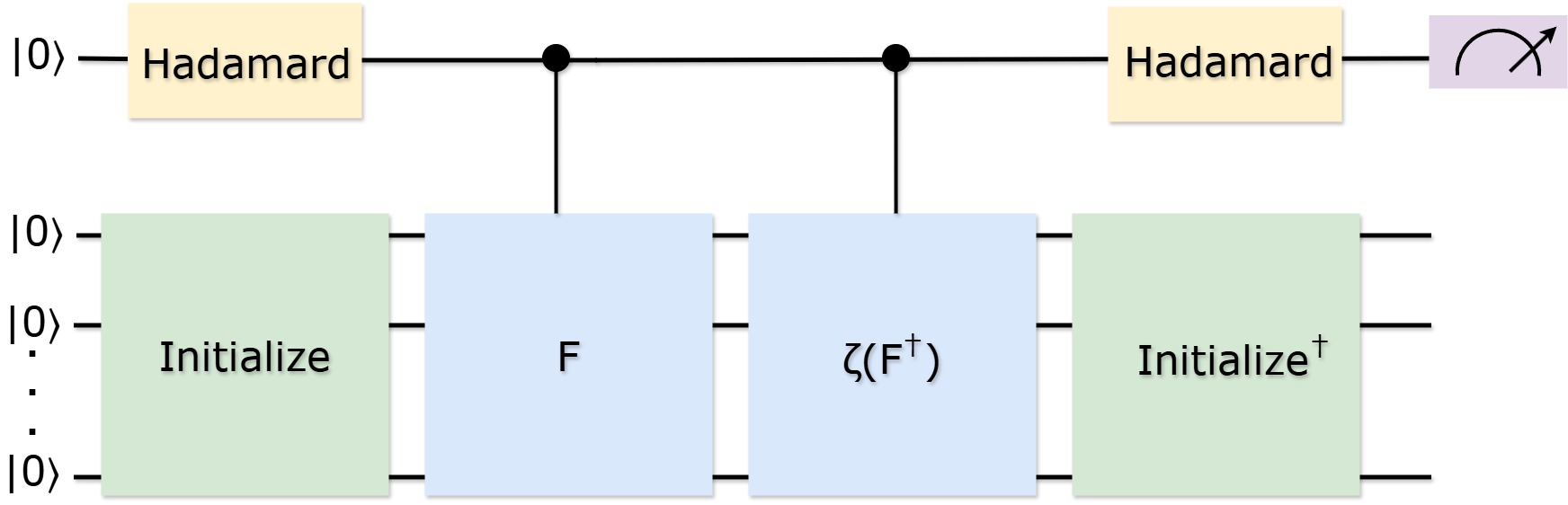}
        \caption{ }
        \label{fig:stage2_top}
    \end{subfigure}

    \begin{subfigure}{\linewidth}
        \centering
        \includegraphics[width=0.90\linewidth]{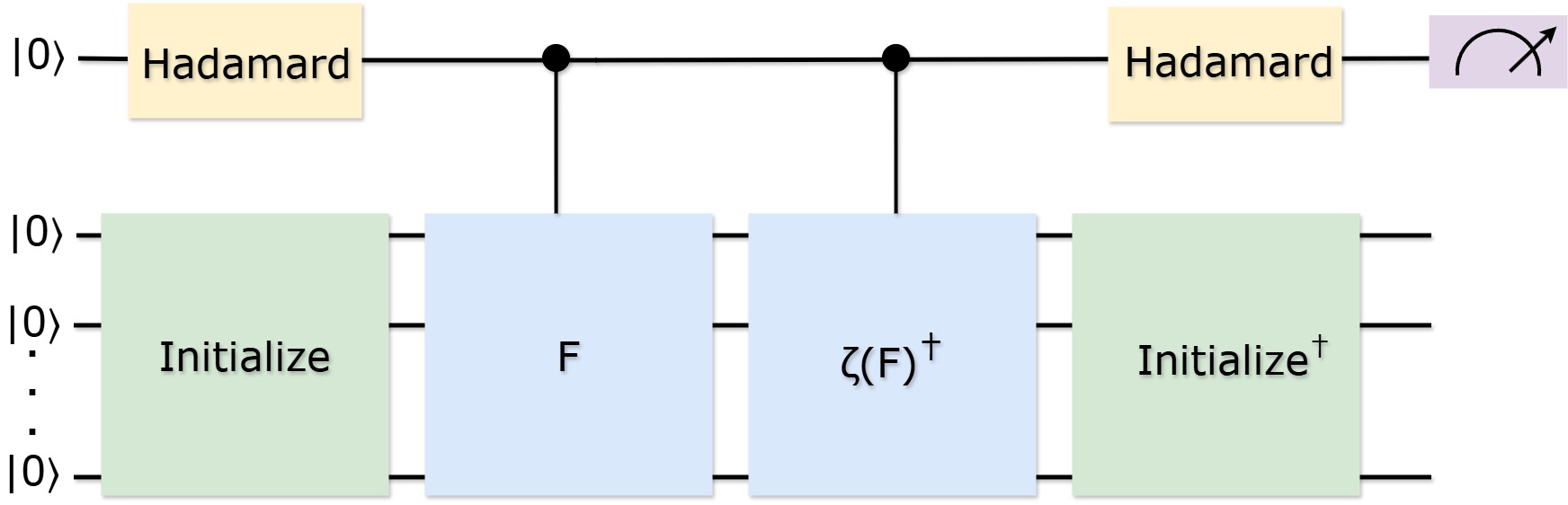}
        \caption{ }
        \label{fig:stage2_middle}
    \end{subfigure}

    \begin{subfigure}{\linewidth}
        \centering
        \includegraphics[width=0.90\linewidth]{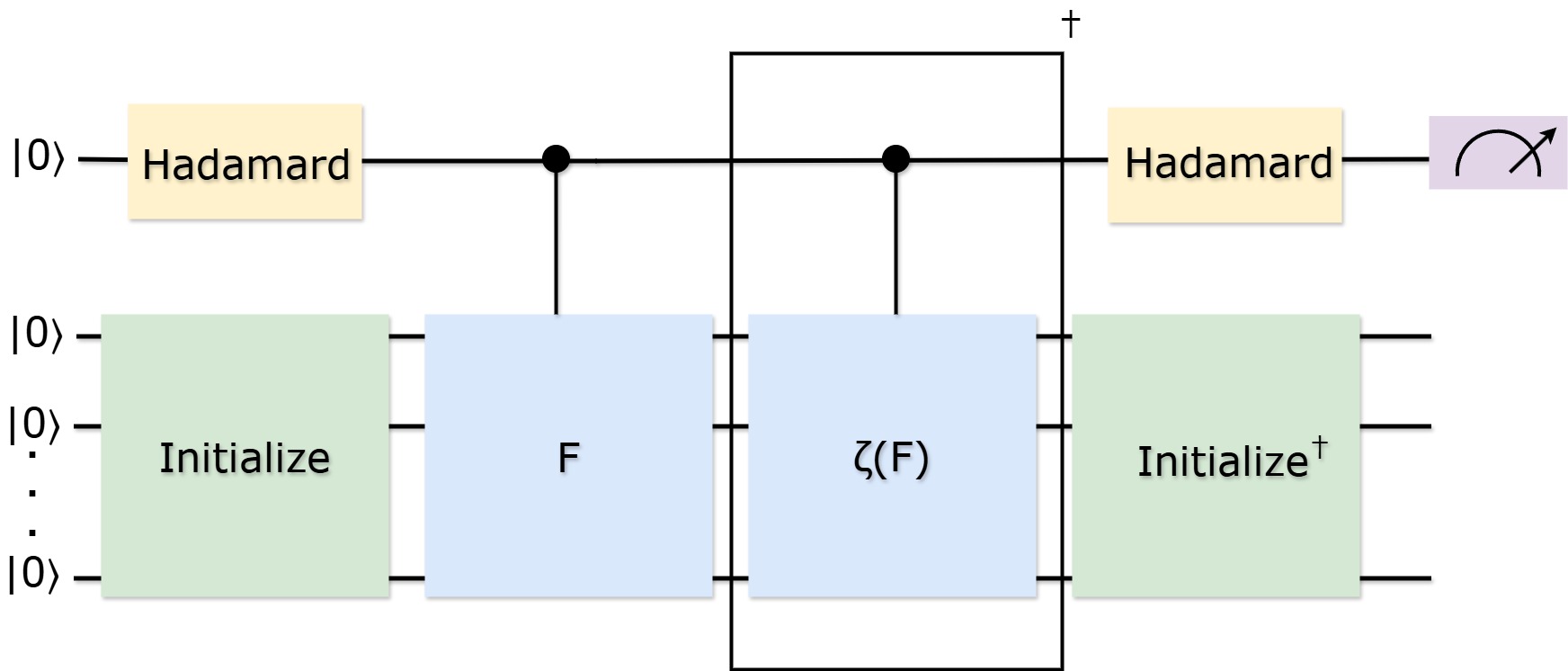}
        \caption{ }
        \label{fig:stage2_bottom}
    \end{subfigure}


    \caption{Stage 2 variant quantum circuits.}
    \label{fig:stage2}
\end{figure}

RetroQ then generates three transformed variants of the original problem, each representing a different way of combining control, adjoint, and the compiler pass operations.
As shown in \autoref{fig:stage2}, the transformed variants are embedded within a Hadamard-test-like verification circuit. This enables the framework to detect semantic differences introduced by compiler passes. 


For each variant, RetroQ records both semantic information, through the expectation value on the auxiliary qubit after measuring it, and structural information. These are then compared against those derived from the baseline circuit. Any deviation from the baseline circuit results indicates either a bug in the compiler pass implementation or an unintended propagation of the pass's effects onto the auxiliary qubit. This provides insight into how the compiler pass should be applied with controlled operators and whether additional user-side modifications are required to properly account for this behavior.

 \subsection{Example Application of RetroQ }
\label{subsec:example_application}
\begin{lstlisting}[language=Python,  caption={RetroQ Execution}, label={lst:stage1}]
import pennylane as qp
from retro_q import RetroQ, Expectation

# 1. Define the target program (F)
def target_function():
    qp.Hadamard(wires=0)
    qp.Hadamard(wires=0)

# 2. Define user expectations for the compiler pass
# We expect the gate count to decrease
exp = Expectation(gate_count='decrease')

# 3. Initialize RetroQ with the pass to test
rq = RetroQ(target_function, compiler_pass=qp.transforms.cancel_inverses, expectation=exp, framework='pennylane')
results = rq.run()

\end{lstlisting}

Consider the PennyLane program in \autoref{lst:stage1} wherein a user defines $F$, \texttt{target\_function}, consisting of two consecutive Hadamard gates on qubit 0. The user aims to verify the \texttt{cancel\_inverses} compiler pass, which is designed to identify and remove redundant adjoint or self-inverse operations, reducing circuit depth. 

In RetroQ, predicates for pass behavior are specified using an \texttt{Expectation} object. Here, the example uses gate count, specifically \texttt{gate\_count=`decrease'} (we expect the compiler pass to successfully identify the redundant gates and remove them). RetroQ currently supports only a gate count predicate, with options for increase, decrease, or unchanged.

\begin{figure}[ht]
         {\includegraphics[  width=0.40\textwidth]{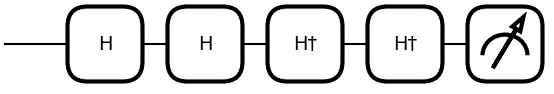}}
        \caption{Stage 1 baseline quantum circuit generated for~\autoref{lst:stage1} when the $Initialize$ operator is the identity operator.}
        \label{fig:listing_stage1}
\end{figure}

In Stage 1, RetroQ constructs a baseline quantum circuit to establish reference structural properties such as total gate count and target expectation values.  For the example in~\autoref{lst:stage1}, the resulting reference circuit is shown in~\autoref{fig:listing_stage1}.


RetroQ then constructs the two circuits defined in~\autoref{fig:stage1} to evaluate the specified compiler pass. For a correctly implemented \texttt{cancel\_inverses} pass with the $Initialize$ operator as the Identity operator, the gate count in these circuits should reduce to 2 (the two Hadamards resulting from execution of $F$ are kept), and the resulting state at the end of the circuits should be $\ket{0}$. 

\begin{figure}[ht]
         {\includegraphics[width=0.5\textwidth]{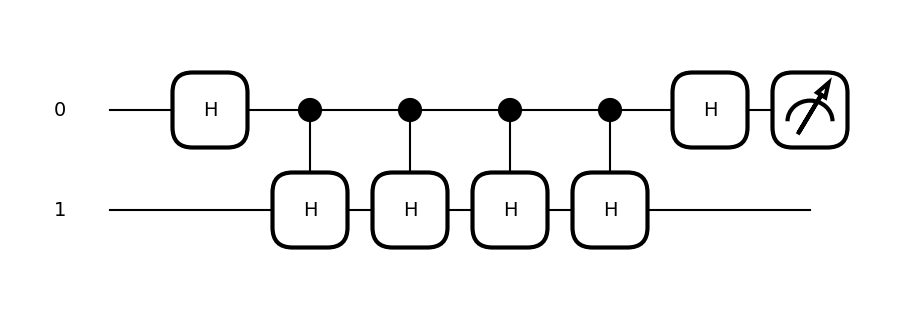}}
        \caption{Stage 2 baseline quantum circuit generated for \autoref{lst:stage1} when the $Initialize$ operator is the identity operator.}
        \label{fig:listing_stage2_baseline}
\end{figure}

In Stage 2, RetroQ will generate the baseline quantum circuit as shown in~\autoref{fig:stage2_baseline} to extract the total gate count and target expectation values. The quantum circuit for our example can be seen in~\autoref{fig:listing_stage2_baseline}.
RetroQ then constructs the three circuits defined in~\autoref{fig:stage2} to evaluate the specified compiler pass. For a correctly implemented \texttt{cancel\_inverses} pass with the $Initialize$ operator as the Identity, the gate count in these circuits should reduce to 4, and the expectation value should be +1.

Any deviations from the expected results will be captured by failure reports that associate each violation with the corresponding circuit description. For the Python program in \autoref{lst:stage1}, the following failure report will be output:
\begin{lstlisting}[
]{python}
{'stage_1_differences': [{
  'detail': 'compiler_pass(adjoint(func))',
  'differences': {
   'gate_count': {'baseline': 4, 'actual': 4, 'ok': False},
   'expectation_value': {
    'baseline': [np.float64(0.9999999999999993)],
    'actual': [np.float64(0.9999999999999993)], 'ok': True
   }}}],
 'stage_2_differences': [{
  'detail': 'control(compiler_pass(adjoint(func))',
  'differences': {
   'gate_count': {'baseline': 6, 'actual': 6, 'ok': False},
   'expectation_value': {
    'baseline': np.float64(0.9999999999999996),
    'actual': np.float64(0.9999999999999993), 'ok': True
   }}}]}}
\end{lstlisting}

As shown above,  the \texttt{cancel\_inverses} compiler pass failed to satisfy the structural properties expectations of gate count reduction for the quantum circuits in Figures \ref{fig:stage2_top} and \ref{fig:s1top_image}. Specifically, the compiler pass failed to recognize that two consecutive adjoints of Hadamard gates cancel each other out, since Hadamard is self-adjoint, leaving the actual gate count at the baseline level instead of reducing it. This limitation extends to the following quantum gates in PennyLane: Paulis $X$, $Y$, $Z$, the identity, $CNOT$, and Toffoli. 

 
\section{Results}
\label{sec:res}
To answer \textbf{RQ3} (What types of compiler transformation errors are
not captured by existing testing and verification approaches?), we evaluated 5 compiler passes using RetroQ: 2 from Qiskit-2.4.1 (\texttt{CommutativeInverseCancellation} and  \texttt{RemoveIdentityEquivalent}), and 3 from PennyLane-0.45.0 (\texttt{cancel\_inverses}, \texttt{single\_qubit\_fusion} and \texttt{pattern\_matching\_optimization}). We chose these compiler passes because they are designed to reduce overall gate count, a structural property RetroQ checks. 

Our evaluation revealed several anomalies that were not captured before, as well as reproduced some bugs that have already been reported. The evaluation consisted of constructing small target quantum programs designed to exercise the intended behavior of each compiler pass (e.g., inverse cancellation, identity removal, gate fusion, and pattern matching) and verifying whether the transformed circuits satisfied the user-specified semantic and structural expectations. For all the bugs identified, we used the Identity operator as the $Initialize$ function. This process identified or replicated 6 bugs (including the one described in the previous section), as detailed below.

\subsection{Qiskit Evaluation: Replication of Existing Bugs}
\subsubsection{Incorrect Commutation Inverse Cancellation with SWAP and Controlled-RZ Gates}

\begin{lstlisting}[language=Python,  caption={Qiskit - Cancelled Swaps~\cite{qiskit_issue_16164}}, label={lst:bugsqiskit1}]
from qiskit import QuantumCircuit
from qiskit.transpiler.passes import CommutativeInverseCancellation

def target_function():
    qc = QuantumCircuit(2)
    qc.swap(0, 1)
    qc.crz(3.2, 0, 1)
    qc.swap(0, 1)
    return qc
    
exp = Expectation(gate_count='unchanged')

rq = RetroQ(target_function, compiler_pass=CommutativeInverseCancellation, expectation=exp, framework='qiskit')
results = rq.run()

\end{lstlisting}
We first evaluate Qiskit's  \texttt{CommuteInverseCancellation} pass by reproducing the previously reported bug in Qiskit Issue \#16164~\cite{qiskit_issue_16164}. The program in ~\autoref{lst:bugsqiskit1} is taken from this bug report because it isolates the commutation logic used by the pass. The circuit places two Swap gates around a Controlled-$RZ$ gate, such that the Swap gates can only be canceled if they are incorrectly commuted through the controlled rotation. 

Here, we expect the number of operations to remain unchanged, i.e., $g(F') = g(F)$. However, in both Stages 1 and 2, the pass incorrectly pushes one Swap gate through the Controlled-$RZ$ gate and then cancels the two Swap gates.

\subsubsection{Zero Angled Parameterized Gate Not Removed}
\begin{lstlisting}[language=Python,  caption={Qiskit - Zero Angle}, label={lst:bugsqiskit2}]
from qiskit.transpiler.passes import RemoveIdentityEquivalent
from qiskit.circuit.parameter import Parameter

def target_function():
  d = Parameter('d')
  qc = QuantumCircuit(2, 2)
  qc.ry(d * 0, 0)
  return qc

exp = Expectation(gate_count='reduce')

rq = RetroQ(target_function, compiler_pass=RemoveIdentityEquivalent, expectation=exp, framework='qiskit')
results = rq.run()

\end{lstlisting}
We next evaluate the \texttt{RemoveIdentityEquivalent} pass, which is intended to eliminate gates whose effects are close to the Identity (up to a global phase and a specified tolerance). This experiment is motivated by Qiskit Issue \#6468~\cite{qiskit_issue_6468}.

Consider the Python program in~\autoref{lst:bugsqiskit2}. Here, $F$ consists of a \texttt{ry(d * 0)} gate, where \texttt{d} is a symbolic parameter. Although the expression is symbolic, it is algebraically equivalent to zero for any value of \texttt{d}. Therefore, the resulting rotation has no physical effect on the circuit. The program is designed to test whether the compiler pass can correctly understand this, rather than relying only on explicitly constant parameters.

 We expect the compiler pass to reduce the gate count ($g(F') < g(F)$) for both Stages 1 and 2. However, no reduction is observed. This issue also affects the $RZ$ and $RX$ gates. This behavior highlights a limitation in the identity detection logic, specifically in its handling of symbolic parameter expressions.



\subsection{PennyLane: Previously-Undetected Issues}
\subsubsection{Crash in Program with Cancel Inverses}
\label{subsubsec:cancel_inverses_bug}
\begin{lstlisting}[language=Python,  caption={PennyLane - Unitary Crash}, label={lst:p1}]
import numpy as np

def target_function():
    U = 1 / np.sqrt(2) * np.array([[1, 1], [1, -1]])
    qp.QubitUnitary(U, 0)
    qp.adjoint(qp.QubitUnitary)(U, 0)
    
exp = Expectation(gate_count='decrease')

rq = RetroQ(target_function, compiler_pass=qp.transforms.cancel_inverses, expectation=exp, framework='pennylane')

results = rq.run()

\end{lstlisting}
We first evaluate PennyLane's \texttt{cancel\_inverses} compiler pass. This pass should remove operations and their adjacent inverses, even when expressed as general unitary matrices (\texttt{qp.QubitUnitary}) rather than built-in quantum gates.

We construct a quantum circuit in \autoref{lst:p1} where a unitary is followed by its adjoint. The expected behavior is that the two operations should cancel, reducing the gate count ($g(F') < g(F)$). However, for both Stages 1 and 2, execution fails at runtime when \texttt{cancel\_inverses} is used, and the following error is thrown: \texttt{ValueError: The truth value of an array with more than one element is ambiguous. Use a.any() or a.all()}. Specifically, the program crashes because the compiler pass attempts to evaluate equality between gate parameters inside the inverse-detection logic.

\subsubsection{Global Phase and Controlled Operator}
\label{subsection:globalphase}
\begin{lstlisting}[language=Python,  caption={PennyLane - Global Phase}, label={lst:p4}]
def target_function():
    qp.Hadamard(wires=0)
    qp.RZ(0.1, wires=0)

exp = Expectation(gate_count='unchanged')

rq = RetroQ(target_function, compiler_pass=qp.transforms.single_qubit_fusion,  compiler_pass_kwargs={'exclude_gates':['Hadamard']}, expectation=exp, framework='pennylane')
results = rq.run()
\end{lstlisting}
Consider the Python program in~\autoref{lst:p4}. Here, $F$ consists of the gates \texttt{Hadamard} and \texttt{RZ} on qubit 0. $F$ was specifically constructed using \texttt{Hadamard} and \texttt{RZ} gates to verify the reliability of the \texttt{exclude\_gates} parameter, which should prevent specific gates from being fused. By placing \texttt{Hadamard} on the exclusion list, the expected behavior was for the compiler to leave the circuit unchanged ($g(F') = g(F)$).
However, after applying the compiler pass to the adjoint of $F$, the operations unexpectedly fused, producing a  \texttt{qp.GlobalPhase} operation, even though Hadamard is self-adjoint. 

This anomaly persisted and intensified in Stage 2. The  \texttt{qp.GlobalPhase}, combined with the controlled operator in~\autoref{fig:stage2}, led to the expectation value of qubit 0 becoming $-1$ rather than $+1$. This highlights the importance of considering global phase in a circuit: global phase outputs from this compiler pass applied on an adjoint of $F$ must be explicitly managed or ignored if the resulting circuit is nested within a control context.

\subsubsection{Adjoint of Quantum Operation Not Considered in Pattern Matching Optimization}
\label{subsection:pmo}
\begin{lstlisting}[language=Python,  caption={PennyLane - Pattern Matching Optimization}, label={lst:pmo}]
def target_function():
    qp.Hadamard(0)
    qp.PauliZ(0)
    qp.Hadamard(0)
    qp.RZ(0.1, wires=0)

exp = Expectation(gate_count='decrease')

ops = [qp.Hadamard(0), qp.PauliZ(0), qp.Hadamard(0), qp.PauliX(0)]
pattern = qp.tape.QuantumTape(ops)

rq = RetroQ(target_function, compiler_pass=qp.transforms.pattern_matching_optimization, compiler_pass_kwargs={'pattern_tapes':[pattern]}, expectation=exp, framework='pennylane')
results = rq.run()


\end{lstlisting}

In ~\autoref{lst:pmo}, the  \texttt{pattern\_matching\_optimization} compiler pass is applied to identify patterns of gates and replace them with more efficient, logically equivalent ones (here, $HZH = X$). In this test, $F$ applies a sequence of self-adjoint gates to test whether the compiler pass could recognize that the adjoint of the sequence is equivalent to the original pattern. For this $F$ and the compiler pass, we expect  $g(F') < g(F)$. 

In both Stage 1 and Stage 2, a critical structural limitation emerged when the compiler pass was applied to the adjoint of $F$. Similar to the example in \autoref{subsec:example_application}, the compiler failed to recognize the sequence $H^{\dagger}Z^{\dagger}H^{\dagger}$
as a match for the $HZH$ segment of the defined pattern, even though these gates are self-adjoint. As a result, the gate counts did not decrease.

\subsection{Discussions and Summary of Findings}
The results for all identified bugs are summarized in \autoref{tab:bug_table}. 
Stage 1 was successful in identifying failures where compiler passes failed to reduce gate counts as expected. These include Qiskit’s failure to remove symbolic zero-angle gates, PennyLane’s incomplete identity recognition for basic Pauli gates and runtime crashes related to how PennyLane's compiler pass logic handles general unitary matrices. 

Stage 2 revealed a risk in PennyLane’s handling of global phase by showing that when circuits are nested in a controlled context, those phases can lead to incorrect expectation values (e.g., $-1$ instead of $+1$). Developers must account for global phases introduced due to compiler passes, as these can become physically significant local phases when a circuit is nested in a controlled operation (which is a common quantum bug~\cite{quetschlich2025experience}).

For the PennyLane bugs, we contacted the PennyLane team to discuss the identified issues and received feedback. The bug described in \ref{subsubsec:cancel_inverses_bug} was confirmed and reported on GitHub~\cite{khan2026pennylane10201comment}. The global phase bug \ref{subsection:globalphase} has been forwarded as a feature request for an optional configuration to preserve phase accuracy in transforms.
Finally, the issue in \ref{subsection:pmo} can be resolved using the \texttt{qp.simplify} transform, which simplifies a QNode by reducing its arithmetic depth or number of rotation parameters~\cite{pennylane_simplify}.

The research answers \textbf{RQ3} by identifying five categories of compiler transformation errors that evade existing testing approaches. These include: (1) symbolic reasoning limitation, where compilers fail to recognize algebraically equivalent operations; (2) abstraction gap, where processing general unitary matrices instead of standard gates causes runtime failures; (3) structural logic error, exemplified by Qiskit's \texttt{CommutativeInverseCancellation} pass incorrectly commuting non-commuting Swap and Controlled-$RZ$ gates to force an invalid inverse cancellation; (4) global phase management error, where a seemingly irrelevant global phase becomes a significant local phase in controlled operations; and (5) adjoint logic failure, where pattern-matching optimizations fail to recognize inverted sequences that are mathematically equivalent to defined patterns. Symbolic reasoning limitation, abstraction gap, and structural logic errors were caught across all circuits in Stages 1 and 2; the global phase management and adjoint logic failure were identified by circuits in Stages 1a and 2a, confirming that each stage within the framework provides essential coverage for distinct types of compiler defects.

Despite these findings, manual test circuit selection may miss critical edge cases. The tested circuits are also relatively small. To address this, a hybrid approach combining compiler pass details with descriptive metadata could be devised for systematic circuit generation. Such an enhancement would enable RetroQ to test a wider variety of passes, including on programs that are too large to run on simulators. Given that the stages of RetroQ are backend agnostic and require only the target function, the inverse of the target function, and the compiler pass under test, RetroQ is suitable for testing compilation passes with quantum hardware. In this setting, tests could be validated based on statistical properties of the output. Such testing at large scales will become increasingly important as hardware continues to advance. 

Another limitation is that our evaluation used only the Identity operator as the initialization function, which may prevent certain bugs from being detected. Consider the example in~\autoref{lst:bugsqiskit1}. Suppose the compiler pass erroneously returned the following sequence for the target function: a Swap gate between qubits 0 and 1, another Swap gate between the same qubits, and a Controlled-$RZ$ gate applied to qubits 0 and 1. Since the gate count remains unchanged and the qubits return to the state $\ket{0}$ after measurement under the identity initialization, the bug would remain undetected. However, if the initialization function instead applied two Hadamard gates to qubits 0 and 1, both qubits would not return to $\ket{0}$, and the bug would be detected.
Devising a strategy for suitable state initialization is a subject for future work.
 


\begin{table}[t]
\centering
\caption{ {Results across Stages 1 and 2 for each listing. A checkmark indicates that the stage caught a bug, and an empty cell indicates the stage did not catch a bug. Each column corresponds to a different stage's circuit instance: 1a and 1b refer to Stage 1 circuits \autoref{fig:s1top_image} and \autoref{fig:s1bottom_image}, respectively, while 2a, 2b and 2c correspond to Stage 2 circuits in \autoref{fig:stage2_top}, \autoref{fig:stage2_middle}, and \autoref{fig:stage2_bottom}.}}

\label{tab:bug_table}

\setlength{\tabcolsep}{3pt}
\renewcommand{\arraystretch}{1.15}
\small

\begin{tabular}{|p{1.3cm}|c|c|c|c|c|}
\hline
\textbf{Listing} & \textbf{1a} & 
\textbf{1b} & 
\textbf{2a} & 
\textbf{2b} & 
\textbf{2c} \\
\hline

\autoref{lst:stage1} & \checkmark &  &\checkmark  &  &  \\
\hline

\autoref{lst:bugsqiskit1}& \checkmark & \checkmark & \checkmark & \checkmark & \checkmark \\
\hline

\autoref{lst:bugsqiskit2}& \checkmark & \checkmark & \checkmark & \checkmark & \checkmark \\
\hline

\autoref{lst:p1} & \checkmark & \checkmark & \checkmark & \checkmark & \checkmark \\
\hline

\autoref{lst:p4} & \checkmark &  & \checkmark &  &  \\
\hline

\autoref{lst:pmo}
& \checkmark &  & \checkmark &  &  \\
\hline

\end{tabular}
\end{table}

\section{Conclusion}
\label{sec:conclusion}
In this work, we performed a systematic analysis of unit tests across four major quantum programming frameworks to address \textbf{RQ1}, revealing that current validation practices for compiler passes are primarily dominated by program-content and program-metric assertions. The analysis also highlights that existing test circuits are generally small and shallow, with the majority of them having depth 1-10. The reliance on simple circuits suggests current methodologies may miss bugs that more complex circuits could catch. 

To address these limitations and answer \textbf{RQ2}, we introduce RetroQ, a black-box framework that automates the verification of quantum compiler passes through retromorphic testing, principles from the Hadamard test, and knowledge of the expected behavior of the compiler pass being tested. To the best of our knowledge, RetroQ is the first framework to adapt retromorphic testing to test quantum compiler passes. By utilizing a dual-program structure, pairing a forward program with its inverse, RetroQ creates an internal verification oracle that eliminates the need for manual derivation of expected outputs or knowledge of internal program structures. 

We answer \textbf{RQ3} by performing experimental evaluations using RetroQ. We successfully reproduced known bugs and uncovered new defects, including an abstraction gap in processing general unitary matrices and an adjoint logic failure.
Furthermore, we identify global phases introduced by compiler passes that lead to qubit expectation value sign inversions when a compiled circuit is applied as a controlled operation. These results demonstrate that the retromorphic testing approach employed by RetroQ is effective in a quantum context. Ultimately, the adoption of specialized, pass-specific testing methodologies like RetroQ will support the reliability and correctness of the next generation of quantum compilers.  Although our current evaluation uses small programs, the RetroQ methodology can generalize to larger programs by automatically constructing their inverses and applying the compiler pass under test without changing the underlying verification procedure.

While RetroQ identified bugs, several areas of future research remain open, including developing procedures to find suitable initial states, generation of test programs, automated derivation of structural expectations for compiler passes, and extension to support additional quantum computing frameworks, such as Cirq and pytket.


\section{Acknowledgments}

MK acknowledges funding from the Natural Sciences and Engineering Research Council of Canada (NSERC) CREATE in Quantum Computing Program, grant number 543245, and the Four Year Doctoral Fellowship. ODM acknowledges funding from NSERC (RGPIN-2022-04609), the Canada Research Chairs program (CRC-2021-00206), and UBC. The authors thank the PennyLane team for feedback on the identified bugs.

\bibliographystyle{IEEEtran}
\bibliography{references}

\end{document}